\documentclass[twocolumn,superscriptaddress,longbibliography,aps,pra,preprintnumbers,showkeys,10pt]{revtex4-2} 

\usepackage{graphicx} % Required for inserting images
\usepackage[utf8]{inputenc}
\usepackage[urlcolor=blue,colorlinks=true,citecolor=blue,linkcolor=blue,pdfstartview={FitH},bookmarks=false]{hyperref}
\usepackage[T1]{fontenc}
\usepackage{xcolor}
\usepackage{amsmath,amssymb,amsfonts}

\newcommand{\size}{0.93\linewidth}
\begin{document}

\preprint{Submitted to: \emph{Acta Physica Polonica A} journal. 
A review article for special issue \emph{40 years of High-$T_c$ SC}}

\title{The Hubbard models with short-range attraction and their generalizations:\\ applications to high-temperature superconductivity and beyond}
\author{Konrad Jerzy Kapcia}
\email[\mbox{corresponding author, e-mail: }]{konrad.kapcia@amu.edu.pl}
\homepage[\mbox{ORCID ID}: ]{https://orcid.org/0000-0001-8842-1886}
%\homepage[\mbox{ORCID ID}: ]{0000-0001-8842-1886}
\affiliation{\mbox{Institute of Spintronics and Quantum Information, Faculty of Physics and Astronomy}, Adam Mickiewicz University in Poznań, Uniwersytetu Poznańskiego 2, PL-61614 Poznań, Poland}

\author{Agnieszka Cichy}
%\email[e-mail: ]{agnieszka.kujawa@amu.edu.pl}
\homepage[\mbox{ORCID ID}: ]{https://orcid.org/0000-0001-5835-9807}
%\homepage[\mbox{ORCID ID}: ]{0000-0001-5835-9807}
\affiliation{\mbox{Institute of Spintronics and Quantum Information, Faculty of Physics and Astronomy}, Adam Mickiewicz University in Poznań, Uniwersytetu Poznańskiego 2, PL-61614 Poznań, Poland}
\affiliation{Institut f\"ur Physik, Johannes Gutenberg-Universit\"at Mainz, Staudingerweg 9, D-55099 Mainz, Germany}

\author{Tomasz Kostyrko}
%\email[e-mail: ]{tkos@amu.edu.pl}
\homepage[\mbox{ORCID ID}: ]{https://orcid.org/0000-0001-7849-0060}
%\homepage[\mbox{ORCID ID}: ]{0000-0001-7849-0060}
\affiliation{\mbox{Institute of Spintronics and Quantum Information, Faculty of Physics and Astronomy}, Adam Mickiewicz University in Poznań, Uniwersytetu Poznańskiego 2, PL-61614 Poznań, Poland}

\date{{\today}}

\begin{abstract}
The attractive Hubbard model and its extensions provide a fundamental framework for understanding superconductivity in systems with short coherence lengths and strong pairing correlations. 
As a minimal model of local electron pairing, the negative-$U$ Hubbard model describes {a} continuous crossover from weak-coupling {Bardeen--Cooper--Schrieffer-like} (BCS-like) superconductivity to the Bose-Einstein condensation (BEC) of tightly bound pairs. 
In this review, we summarize the theoretical foundations and major developments of attractive Hubbard-type models, emphasizing the contributions of Roman Micnas and Stanis{\l}aw Robaszkiewicz, and their collaborators. 
We discuss local-pair superconductivity, the BCS-BEC crossover, and representative theoretical approaches, including {the} mean-field Hartree-Fock approach, random-phase approximation,  and slave-boson methods. 
Connections to cuprate superconductors and ultracold atomic gases are also highlighted. 
{Taken together,} these developments establish the attractive Hubbard model as a versatile paradigm for describing superconductivity and collective quantum phenomena in strongly interacting fermionic systems.
\end{abstract}


\maketitle

\section{Introduction}
\label{sec:intro}

The discovery of high-temperature superconductivity in copper-based ceramics was one of the most important scientific {events} of the twentieth century. 
In the first system from the family of copper oxide perovskite, i.e., La-Ba-Cu-O system, superconductivity was reported below {a} critical temperature of \mbox{$T_c \approx 30$}~K \cite{BednorzZPhysB1986}. 
Subsequently, superconductivity observed at temperatures above the boiling point of liquid nitrogen was found, e.g., in {the} Y-Ba-Cu-O system (\mbox{$T_c \approx 92$}~K) \cite{WuPRL1987} and in {the} Hg-Ba-Ca-Cu-O system (\mbox{$T_c \approx 135$}~K; with the record one \mbox{$T_c \approx 164$}~K under pressure of $31$~GPa \cite{GaoPRB1994}).

These materials have unique features {that} are distinct from {those} known before (for extended review see e.g., Refs. \cite{BednorzRMP1988, PickettPRP1989, MicnasRMP1990, MicnasRev1996, MicnasNATO1997, MicnasCMP1998, Keimer2015, BozovicPhysA2019, Zhou2021}). 
{Typically,} they are narrow-band systems {in which} electron correlations are of {great} importance \cite{PaschenNRP2021}.
{Their} superconductivity is strongly anisotropic and has a (quasi-) two-dimensional nature originating from {the} Cu-O layers and the carriers in the CuO$_{2}$ planes are holes in the doped Mott insulator.
In normal state, {these materials} are poor conductors and exhibit a linear temperature dependence of the resistivity.
There is also a low or negligible isotope effect.
{The} superconducting coherence length $\xi$ is very small {in these materials} (\mbox{$\xi_{ab} \approx 1-3$} nm, \mbox{$\xi_{c} \approx 0.1-0.5$} nm), and its anisotropy reflects their strongly anisotropic layered structure \cite{Wesche2025}.
Together with very large upper critical {field $H_{c2}$ and a positive temperature slope of $H_{c2}$} near $T_{c}$,
the systems are classified as extreme second-type superconductors.

The microscopic BCS theory of superconductivity {(named after Bardeen, Cooper, Schrieffer)} describes the instability of {a} Fermi liquid toward the formation of bound electron pairs due to effective attraction \cite{BardeenCooperSchrieffer1957A, BardeenCooperSchrieffer1957, FrohlichPR1950, CooperPR1956, SchriefferRMP1999}.
Electrons near the Fermi surface (in relatively narrow energy range associated to {the characteristic} energy of lattice excitation $\hbar \omega_{D}$; corresponding to {the} superconducting energy gap $\Delta$) and with opposite momenta and spins form the Cooper pairs.
These boson-like pairs are highly delocalized with a characteristic size  much larger than the lattice spacing and the average inter-particle distance ($\xi_{BCS} \approx 10^2 - 10^3$ nm). 
Thus, the Cooper pairs strongly overlap.
In this theory, {the} pair formation and {the} establishment of long-range superconducting phase coherence occur simultaneously at {the} transition temperature $T_{c}$.
Despite {the} strong direct repulsion between electrons, {a} relatively weak (effective) attraction binding the pairs originates {from} the coupling of electronic system with the lattice vibrations. 
Such a mechanism provides the limit for {the} critical temperature {of} \mbox{$T_{c} \approx 30$}~K \cite{CohenAIP1972}; however, {the} MgB$_2$ compound is a BCS-like superconductor with 
\mbox{$T_{c} \approx 39$}~K, {in which} the electron-phonon mechanism is optimized due to an unusually favorable combination of high-frequency phonons, strong electron-phonon coupling, and a multiband electronic structure~\cite{NagamatsuNature2001, MazinPRL2002, PuzniakPhysC2007}.

Although the BCS theory successfully explains conventional low-temperature superconductors (like, for example, pure metallic elements, intermetallic alloys, and Chevrel's phases), many narrow-band systems, (in particular {the} copper-oxide systems described at the beginning) exhibit features {that} do not fit a group of ordinary {electron-phonon} BCS superconductors and {therefore} require a description within a different theory. 
Despite almost four decades of intensive research, {any} universally accepted microscopic theory of high-$T_{c}$ superconductivity has not yet emerged.
Numerous theoretical scenarios have been proposed, including, for example, resonating-valence-bond \cite{AndersonScience1987, AndersonPRL1987, DallaNatPhys2015}, spin-fluctuation \cite{ScalapinoPRB1986, MonthouxPRL1991, MoriyaAP2000, ScalapinoRMP2012}, electron-phonon and polaronic \cite{AlexandrovRPP1994, LanzaraNature2001, GarciaACMP2010}, and quantum-critical \cite{SachdevRMP2003, Keimer2015, PaschenNRP2021} approaches, as well as hole superconductivity \cite{HirschPLA1989, HirschPRL2001, Hirsch2024} and local-pairing theories.

In the present context, we focus on local-pairing theories, in which the essential objects are preformed electron pairs existing above $T_{c}$.
Superconductivity emerges through {the} phase coherence of these pairs rather than through pair formation at {a characteristic temperature} $T_{p}$.
These phenomena can be described microscopically in the framework of different effective models such as {the} boson-fermion model \cite{RanningerPhysBC1985, FriedbergPLA1989, FriedbergPRB1989, MicnasPRB2002, DomanskiPRL2003, DomanskiPRB2003, MicnasPhysC2003, PawlowskiJS2004, MicnasSB2005, Micnas2007, KrzyszczakJPCM2010,Pawlowski2010} or extended Hubbard models {(EHMs)} with various attractive interactions (for early reviews see, e.g., Ref.~\cite{MicnasRMP1990, MicnasRev1996, MicnasNATO1997, MicnasCMP1998}). 
This {theoretical} framework is closely related to the BCS-BEC crossover {phenomenon \cite{ChenRMP2024},} which will be discussed later in this work.
Suggestions that this class of microscopic models can be used for the description of superconductivity in the high-temperature superconductors were {made} shortly after its discovery in cuprates (see, e.g., \cite{EmeryPRL1987, ZhangPRB1988, MicnasJPC1988, MicnasPRB1988, MicnasPRB1989, MicnasRMP1990}) and {were} motivated by coherence lengths comparable to only a few lattice spacings, suggesting that pairing correlations may {indeed become} local. 
However, one should underline that these theories were originally developed before the discovery {of high-temperature superconductivity} (in particular, those based on the negative-$U$ (extended) Hubbard models \cite{ChaoPRB1979, RobaszkiewiczPRB1981C, RobaszkiewiczPRB1981D, RobaszkiewiczPRB1981A, RobaszkiewiczPRB1981B}) and later applied {in} attempts {to} descriptions of high-$T_{c}$ superconductivity  in real systems.

This work provides a review and discussion  of the extended Hubbard models with various extensions.
It particularly focuses on the works developed over several decades by the Pozna{\'n} school at Adam Mickiewicz University {(AMU)} conducted by Professors Roman Micnas and Stanis{\l}aw Robaszkiewicz (who were heads of former {the} Department of Solid State and {the} Department of Electronic States of Solids, respectively).
{This paper} also demonstrates how theoretical concepts initially introduced in the context of superconductivity have been relevant to {the} interpretation of experimental data in unconventional superconductors and  a much wider range of quantum many-body systems.

The remainder of this work is organized as follows.
Section \ref{sec:EHMgeneral} is devoted to a discussion of superconductivity in the framework of {an} extended Hubbard model with intersite density-density interactions with on-site pairing (\mbox{$U<0$}, Sect. \ref{sec:AHMonsite}) and intersite pairing ($W<0$, Sect. \ref{sec:EHMintersite}), together with a review of effects of other-type intersite interactions (Sect. \ref{sec:other}).
Section \ref{sec:realsystems} contains {a} discussion of an experimental scope of the theory in the context of solid-state systems (Sect. \ref{sec:solidstate}) as well as ultracold atoms on optical lattices (Sect. \ref{sec:ultracold}).
Finally, Sect. \ref{sec:summary} {presents the} summary and final remarks.

\section{Extended Hubbard model and superconductivity}
\label{sec:EHMgeneral}

The extended Hubbard model constitutes one of the most fundamental microscopic frameworks for investigating the interplay between superconductivity, charge ordering, and phase separation in strongly correlated electron systems. 
Its Hamiltonian has the following form:
\begin{eqnarray}
\label{eq:EHM}
\hat{H}& = &  \sum_{ i,j, \sigma} t_{ij} \left( \hat{c}_{i\sigma}^{\dagger} \hat{c}_{j\sigma} + \hat{c}_{j\sigma}^{\dagger} \hat{c}_{i\sigma} \right) - \mu \sum_{i} \hat{n}_{i} \\
& + & U \sum_{i} \hat{n}_{i\uparrow} \hat{n}_{i\downarrow}+\frac{1}{2} \sum_{i,j} W_{ij} \hat{n}_{i}\hat{n}_{j} , \nonumber
\end{eqnarray}
where $\hat{c}^{\dagger}_{i\sigma}$ ($\hat{c}_{i\sigma}$) is a creation (annihilation) operator of a fermionic particle (e.g., an electron) with spin \mbox{$\sigma \in \{ \uparrow,\downarrow \}$} at lattice site $i$, which fulfills the anticommutation relations: 
\mbox{$\{ \hat{c}_{i\sigma}, \hat{c}^{\dagger}_{j\rho} \} = \delta_{ij}\delta_{\sigma \rho}$}; 
$\{ \hat{c}_{i\sigma}, \hat{c}_{j\rho} \} = \{ \hat{c}^\dagger_{i\sigma}, \hat{c}^\dagger_{j\rho} \} = 0$ for any sites $i$, $j$ and spins $\sigma$, $\rho$. 
The number operator is {given by} \mbox{$\hat{n}_{i\sigma} = \hat{c}^{\dagger}_{i\sigma} \hat{c}_{i\sigma}$}, whereas the total occupation of site $i$ is defined by $\hat{n}_{i} = \hat{n}_{i\uparrow} + \hat{n}_{i\downarrow}$.
{The symbol} $t_{ij}$ is the hopping amplitude between sites $i$ and $j$, $U$ denotes the on-site (local) density-density interaction, $W_{ij}$ stands for the intersite density-density interactions.
$\mu$ is the chemical potential, which is associated with the total concentration \mbox{$n \in \left[0,2\right]$} (number of particles per site) by {the} relation \mbox{$n = \left( 1/L \right) \sum_{i} \left\langle \hat{n}_{i} \right\rangle$}, where $L$ is {the} number of lattice sites. 
If it is not stated explicitly, the following discussion is restricted only to the case of nearest-neighbor (NN) hopping and NN intersite interactions (denoted as $t$ and~$W$, {respectively).}
The repulsive Hubbard model (\mbox{$U>0$}, \mbox{$W=0$}) has served as a paradigmatic framework for describing correlation-driven metal-insulator transitions ~\cite{HubbardPRSL1963, PennPR1966, Spalek1987, GeorgesRMP1996, ImadaRMP1998, ZhangPRB2026}.

\emph{Origin of the local attraction.} 
The single-orbital model parameters appearing in (\ref{eq:EHM}) {(as well as} in (\ref{eq:hamHCB}) and (\ref{eq:EHMadd}) introduced later) should be regarded as effective quantities renormalized from their bare values. 
Several microscopic mechanisms can lead to an effective short-range attraction between electrons~\cite{MicnasRMP1990, MicnasCMP1998}.
The most natural one is {the} interplay between electrons and bosonic degrees of freedom.
In particular, electron-lattice coupling  gives rise to the formation of small polarons (i.e., electrons surrounded by local lattice distortion) \cite{Alexandrov1981A, Alexandrov1981B, AlexandrovPRL1986, RobaszkiewiczPRB1987, HanPRL2024}. 
Two such polarons may attract each other and create a small bipolaron \cite{AlexandrovRPP1994}.
It should be emphasized that, in the BCS theory, the attractive interaction arises from the exchange of virtual phonons between two distinct electrons rather than from the dressing of a single electron by a heavy polarization cloud \cite{BardeenCooperSchrieffer1957A, BardeenCooperSchrieffer1957, FrohlichPR1950, CooperPR1956, SchriefferRMP1999}.
Effective attractive interactions can also result from the coupling of electrons to other bosonic excitations, such as plasmons \cite{Takada1978, IntVeld2DMater2026} or excitons \cite{LittlePR1964, AllenderPRB1973}, which are collective modes of electronic origin. 
Furthermore, interactions between electrons and additional electronic subsystems (orbitals) present in solids or molecular complexes may generate an effective attraction in a reduced single-orbital description \cite{VarmaPRL1988, MatsuuraAAPPS2022}. 
The induced attractive interaction can, under appropriate conditions, overcome the bare Coulomb repulsion between particles with charges of the same sign.
Finally, the lattice sites do not necessarily correspond to the actual atomic positions of the underlying system and they may instead represent effective sites arising from a simplified model description of the particular system.

\emph{Attraction-repulsion transformation.}
Consider {the} \mbox{$U<0$} case {in the} model (\ref{eq:EHM}).
Such {a} term favors the formation of intra-site {electron pairs with antiparallel spins.}
Since {the} other terms {in} (\ref{eq:EHM}) do not depend on spins, the model does not exhibit any magnetic order \cite{LiebPRL1989}; in particular, {neither} ferromagnetic nor antiferromagnetic orderings exist.
On the bipartite lattice one can introduce the  canonical transformation {(the} \mbox{$U\leftrightarrow-U$} transformation) \cite{ChaoPRB1979, RobaszkiewiczPRB1981C, MicnasRMP1990} {that} transforms {the} model (\ref{eq:EHM}) into the EHM with on-site repulsion \mbox{$U'=-U$} and intersite Ising-like exchange \mbox{$J'=W$} (ferromagnetic if \mbox{$W<0$}) in {an} external magnetic field \mbox{$h'= \mu-U/2-zW$} ($z$ stands for {the} number of the nearest-neighbors) and single particle per site (i.e., \mbox{$n'=1$}).
In particular, it means that for \mbox{$W=0$} there is a direct correspondence between negative-$U$ and positive-$U$ Hubbard models (both at the half-filling). 
The long-range antiferromagnetic (AF) order occurring for \mbox{$U>0$} corresponds to the superconducting (SC) and charged-ordered (CO) phases for {the} \mbox{$U<0$} case, which are strictly degenerated due to {the} $SU(2)$ symmetry of the model with~\mbox{$U>0$}.

Note also  that the large repulsive limit (i.e., $U\gg |t|$) of {the} model (\ref{eq:EHM}) is another approach for description of unconventional superconductivity.
The resulting microscopic Hamiltonian is called the $t$--$J$ model and  initially it was obtained as a perturbative expansion of the Hubbard model \cite{SpalekPhysBC1977, ChaoJPC1977, SpalekPRB1978} and later, {both this model} and its extensions {have been} also used (together with inclusion of $W$ interaction) successfully in the context of high-$T_{c}$ superconductors (for a review {of the subject} see, e.g., Refs.~\cite{SpalekAPPA2007, SpalekAPPA2012, SpalekPR2022, SpalekAPPA2023, SpalekAPPB2026}).

\subsection{The model with onsite pairing ($U<0$)}
\label{sec:AHMonsite}

{The} model (\ref{eq:EHM}) with on-site attraction ($U<0$) provides a minimal microscopic description of superconductivity in narrow-band systems with short-range attractive interactions. 
In this model, contrary to the BCS theory, the range of electronic energies for which attraction between particles occurs is wider and is of the order of the bandwidth.
The competition between electron pair condensation and charge ordering (controlled by {the} interaction $W$) leads to a variety of ordered phases, including the SC (with on-site $s$-wave pairing) and CO phases, as well as phase-separated (PS) states. 
In the strong-coupling regime, this model reduces to a system of hard-core bosons, establishing a direct connection between BCS superconductivity and Bose-Einstein condensation (BEC) of tightly bound pairs. 

\begin{figure}[t]
    \centering
    \includegraphics[width=\size]{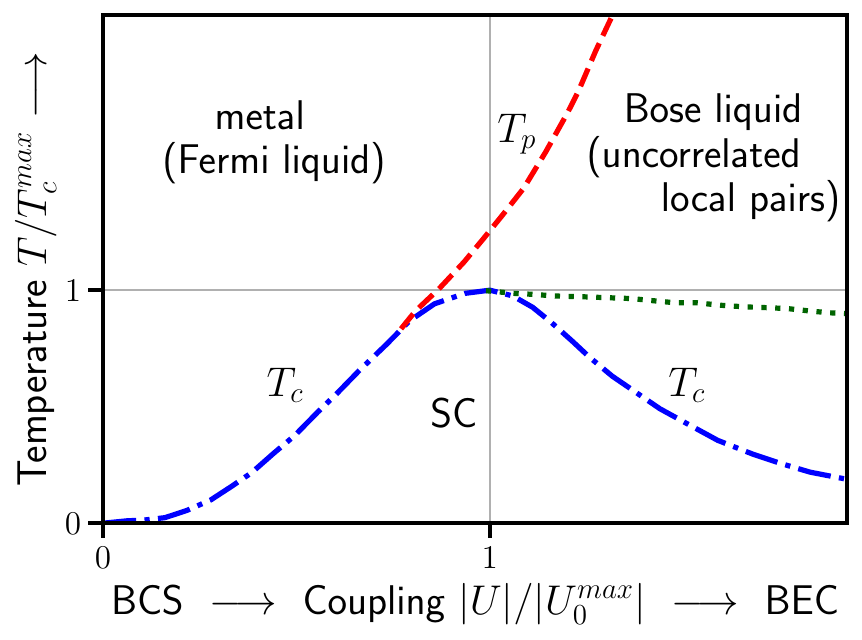}
    \caption{Schematic diagram illustrating the BCS-BEC crossover (between weak-coupling to local-pair regimes).
    The dash-dotted line denotes {the critical temperature} $T_{c}$ for the SC phase, {while}
    $T_{p}$ (dashed line) indicates a characteristic temperature of pair formation.}
    \label{fig:BCSBECcrosover}
\end{figure}

\emph{The BCS-BEC crossover.}
A schematic phase diagram of {the} model (\ref{eq:EHM}) (for $|U| \gg |W| $ and concentrations \mbox{$|n-1| \approx 1$}; the dilute limit) illustrating the BCS--BEC crossover is shown in Fig.~\ref{fig:BCSBECcrosover}. 
The dash-dotted line marks the superconducting transition temperature $T_{c}$, {which separates} the superconducting and normal states. 
In the weak-coupling regime (i.e., when \mbox{$U\ll t, W$}), \mbox{$T_c \sim t \exp\left[-1/\left(|U|N(0)\right)\right]$}, where $N(0)$ is the density of states (DOS) at the Fermi level (similarly as in the BCS theory).
In the strong-coupling limit (\mbox{$U\gg t$}) \mbox{$T_c \sim t^2/|U|$}, reflecting the Bose condensation of tightly bound local pairs. 
The intermediate crossover region, which is the most complex one for the analyses, is characterized by {the} pair formation at a temperature \mbox{$T_p \sim |U|$}. 
For \mbox{$T>T_{p}$}, {the} local pairs are thermally
dissociated.
In this regime, the pair formation and the establishment of long-range phase coherence occur at different energy scales. 
As a consequence, precursor pairing correlations may persist above $T_{c}$, leading to {the emergence of} a pseudogap in the single-particle excitation spectrum and departures from conventional Fermi-liquid behavior. 
The dotted continuation of {the $T_{c}$} line corresponds to the transition line obtained within the continuum model~\cite{NozieresJLTP1985}.

Notice that this general dependence of $T_c$ on $|U|$ for {the} attractive Hubbard model was also found in different dimensions and {theoretical} approaches. 
In particular, in the limit of infinite dimensions (\mbox{$z\rightarrow \infty$}), similar features were found within the dynamical mean-field theory (DMFT) \cite{CaponePRL2002, ToschiPRB2005, PetersPRB2015, SakaiPRB2015} (which provides an exact treatment of all local interactions), including some evidence of pseudogap \cite{PetersPRB2015} or hidden fermionic excitations \cite{SakaiPRB2015}. 
Also the studies with use of {the} $T$-matrix method \cite{MicnasPRB1995A} and the slave-boson mean-field method \cite{BulkaPRB1996, BakJPCM1998} were done (both for \mbox{$W=0$}).
Obviously, in 2D systems, temperature $T_c$ corresponds to {a} Kosterlitz-Thouless (TK) transition, however, even weak interlayer interaction can introduce long-range order \cite{MicnasRMP1990}.
Note that in the presence of the external Zeeman magnetic field, spin-polarized SC phases can also appear \cite{KujawaAPPA2008, KujawaEPL2011, CichyAP2014, CichyAP2015, PtokPRA2017, PtokJSNM2018, CichyPRB2022, CichyJMMM2024}.

As it was mentioned previously, in the half-filling case and for \mbox{$W=0$}, {the SC and CO phases are strictly degenerated due to the $SU(2)$ symmetry of the model}.
The structure of the diagram is similar to those presented in Fig. \ref{fig:BCSBECcrosover}. 
{It features} with two temperature lines $T_{c}$ and $T_{p}$, as well as with the SC/CO region limited by a transition line with maximum $T_{c}$ {occuring at} intermediate values of $|U|/t$ \cite{RobaszkiewiczPRB1981C, MicnasRMP1990, ChaoPRB1979}.
In {the temperature range} \mbox{$T_{c}<T<T_{p}$} (and for larger $|U|/t$), there is a region of short-range order {where} the system is a normal resistive metal with charge $2e$ and strongly suppressed magnetic susceptibility.
{The degeneration} between the SC and CO phases can be removed by \mbox{$W \neq 0$} interaction, which either promotes the SC phase (if \mbox{$W<0$}) or stabilizes the CO phase (if \mbox{$W>0$}).
Also deviation from the \mbox{$n=1$} case favors the SC ordering.
This issue, for arbitrary concentration \mbox{$n \neq 1$}, will be discussed in details for two limits of the model, separately.

\begin{figure}[b!]
    \centering
    \includegraphics[width=\size]{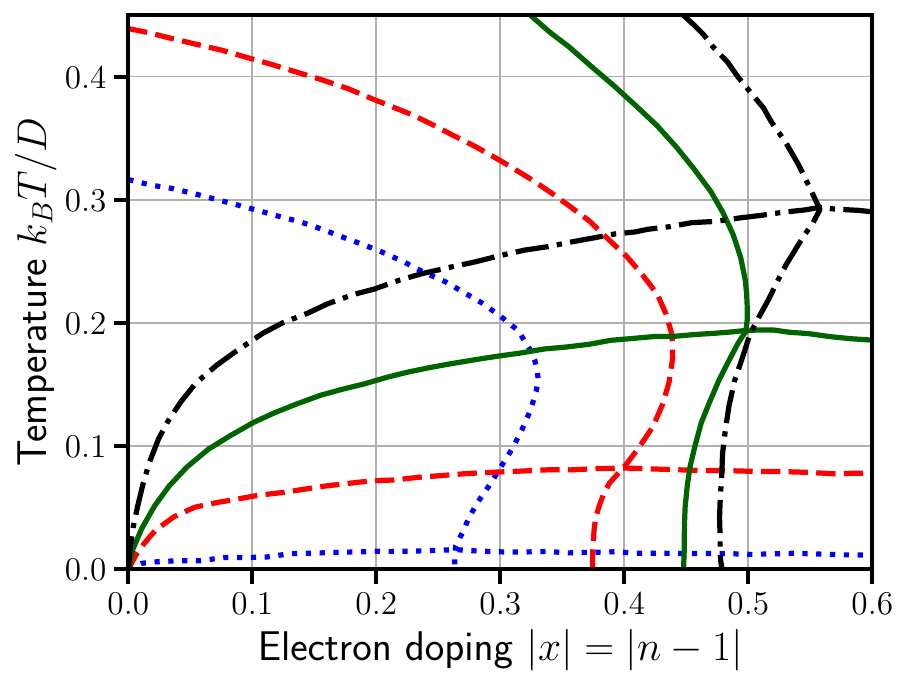}
    \caption{Phase diagram of {the} model (\ref{eq:EHM}) for the square density of states.
    \mbox{$U/D=-0.4,-0.8,-1.2, -1.6$} corresponds to dotted, dashed, solid, and dash-dotted curves (\mbox{$zW/D = 0.55$} fixed).
    Relative positions of {the} four phases (NO, SC, M, CO; clockwise around multicritical point) are the same as in Fig.~\ref{fig:PD-HCB-vsn}.
    Data {taken} from Fig.~1 in~\cite{RobaszkiewiczPRB1981B}.}
    \label{fig:PD-WKL-vsn}
\end{figure}

\subsubsection{The weak-coupling limit of the model}
\label{sec:weakcopuling}

In the weak-coupling limit, the phase diagram of model (\ref{eq:EHM}) with \mbox{$U<0$} was analyzed in \cite{RobaszkiewiczPRB1981A, RobaszkiewiczPRB1981B} within the broken-symmetry Hartree-Fock approximation (HFA). 
At the ground state three distinct phases can exist \cite{RobaszkiewiczPRB1981A}. 
The CO phase exist only at \mbox{$n=1$}. 
For \mbox{$n \neq 1$},  the M phase appears, in which both {types of} orders (i.e., the SC and CO states) coexist. 
At \mbox{$T=0$}, at some critical doping \mbox{$x=n-1$}, there is a continuous transition into a pure SC phase. 
Increasing {the values of} $|U|$ and $W$ extends the M phase region.
{Regardless} of $U$ and $W$, the finite temperature  phase diagram as a function of $|x|$ has the structure consisting of four distinct regions. 
Figure \ref{fig:PD-WKL-vsn} shows a representative diagram. 
Near half-filling (\mbox{$x=0$}), the CO phase is stable, whereas for larger $|x|$, the SC phase occurs up to \mbox{$|x| = 1$}. 
Between these  regions, the M phase exists, which is a result of the competition between the {two} ordered states.
The weight of the SC component in the M phase increases with decreasing temperature, thus the appearance of {features characteristic of the SC phase} is expected near the transition from the metallic CO phase {(for} \mbox{$n \neq 1$}) to the M phase with lowering temperature.
Finally, at high temperature the nonordered (NO) phase is stable, where all orders are absent.
All transition lines (which are continuous ones) merge in the multicritical point.
With increasing temperature {the} following sequences of transitions were predicted ($x \neq 0$): 
(i) SC-NO (large $|x|$), (ii) SC-M-CO-NO (intermediate $|x|$), and (iii) M-CO-NO (small $|x|$).
For {values of} $zW/D$ comparable or greater than $|U|/D$, the sequence SC-NO-CO-NO with increasing $T$ also appears, which is a special feature of the weak-coupling case~\cite{RobaszkiewiczPRB1981B, MicnasRMP1990}.

It is worth noting that {the} Hubbard model with \mbox{$U<0$} (\mbox{$W=0$}), even in the weak-coupling limit, exhibits essential differences from the BCS theory. 
In particular, due to electron pairing in the whole Brillouin zone, the effective half-bandwidth $D$ (instead of $\omega_{D}$) plays a role of the energy scale.
For example, for the square DOS (i.e., \mbox{$\rho(\varepsilon)=1/(2D)$} for \mbox{$|\epsilon|<D$} and \mbox{$\rho(\varepsilon)=0$} otherwise), one gets the following expressions for {the} critical temperature $T_{c}$ and {the} gap value $\Delta_{0}$ at the ground state (after \cite{RobaszkiewiczPRB1981B, MicnasRMP1990}), {respectively,}
\begin{equation}
k_{B} T_{c} =  1.14 D \sqrt{n(2-n)} \times \exp{\left( -2D/|U|\right)},  
\end{equation}
{and}
\begin{equation}
    \Delta_{0}  =  D \sqrt{n(2-n)} \times \left[ \sinh{\left( 2D/|U| \right)} \right]^{-1}.
\end{equation}
Using different shapes of DOS introduces changes in the expressions above, including {the} ratio \mbox{$\Delta_{0} / ( k_{B} T_{c} )$}, which deviates from the BCS value of~$1.76$.

Further studies of {the} negative-$U$ model (\ref{eq:EHM}) including PS states {have shown} that, for \mbox{$W>0$}, the M phase might be not stable and that the CO and SC phases coexist in the PS state {rather than within} the homogeneous M {phase \cite{RobaszkiewiczAPPA1996, PawlowskiAPPA1997}.}
However, {the} inclusion of long-range repulsion beyond next-nearest-neighbor (NNN) {interactions} \cite{RobaszkiewiczAPPA1996}, or {the incorporation of an} additional $\eta$-wave SC order in the M phase \cite{BakAPPA2004} lead to the M phase stabilization.

The random-phase approximation (RPA) is a powerful method {that} goes beyond {the} standard mean-field approach \cite{KostyrkoPRB1992, KostyrkoAPPA1993, KostyrkoPRB1994}. 
Combining the RPA with a diagrammatic method, {an} analytical description of the evolution of {the} collective-excitation spectrum for {the} model (\ref{eq:EHM}) {accros} the entire range of \mbox{$U<0$} (for \mbox{$W \neq 0$}, and arbitrary $n$) was determined in the SC state at \mbox{$T=0$}~\cite{KostyrkoPRB1992, KostyrkoAPPA1993}. 
The collective-mode energies, determined by the poles of the response functions, exhibit a linear dispersion at long wavelengths, $\omega \sim q$. 
The corresponding velocity evolves smoothly  between {the limits of} \mbox{$|U| \ll D$} and \mbox{$|U| \gg D$}. 
Roton-like minima were found in the spectrum near the Brillouin zone boundary for the square lattice.
Furthermore, \mbox{$W>0$} can stabilize CO states {that are} incommensurate with the lattice period for \mbox{$n \neq 1$}, whereas \mbox{$W < 0$} results in instability toward phase separation (electron droplet formation).
Such incommensurate CO states are absent in the bosonic limit {if} only the NN repulsion $W$ is considered \cite{MicnasRev1996}.
Longer-range repulsion also changes the linear dispersions of collective modes (for small $q$ {at} the center of the Brillouin zone) into a square root one, \mbox{$\omega \sim \sqrt{q}$}, for 2D system; whereas in 3D case, {by contrast,} the acoustic branch is transformed into a plasma branch {where $\omega$ is} not dependent on $q$.
In Ref. \cite{KostyrkoPRB1994}, {a} gauge-invariant RPA theory was developed to describe the Meissner effect in a lattice superconductor at \mbox{$T=0$} (model (\ref{eq:EHM}) with \mbox{$U<0$}, \mbox{$W=0$}).
The main result is that the evolution from the Pippard-type behavior (in the weak-$U$ limit) to the London-type behavior (for the strong coupling) is smooth.
{Taken together,} these results provided the analytical descriptions of the evolution of collective excitations across the entire BCS-BEC crossover and constitute an important element of the physical picture of superconductivity {within the} negative-$U$ Hubbard model (cf. also {the} $T$-matrix method results in~\cite{MicnasPRB1995A}).

\begin{figure}
    \centering
    \includegraphics[width=\size]{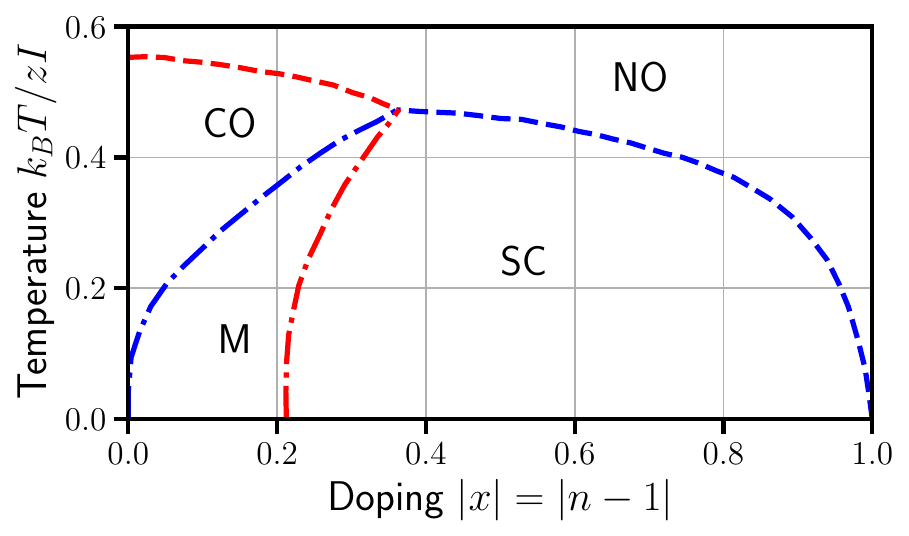}
    \caption{Phase diagram of {the} model (\ref{eq:hamHCB}) for \mbox{$K/I=1.1$} (the MFA result).
    The SC and CO states coexist in the homogeneous M phase.
    All transitions are continuous.
    Depending on {the} other model parameters, the M phase may be unstable toward the PS state with coexisting SC and CO domains.
    Data {taken} from Fig.~4 in~\cite{RobaszkiewiczPRB1981C}.}
    \label{fig:PD-HCB-vsn}
\end{figure}

\subsubsection{The strong-coupling limit of the model}
\label{sec:strongcoupling}

Using degenerate perturbation theory, in the limit $|U| \gg t_{ij}$, {the} model (\ref{eq:EHM}) can be written as the effective pseudospin Hamiltonian, which is equivalent {to} charged hard-core boson model \cite{ChaoPRB1979, RobaszkiewiczPRB1981C, RobaszkiewiczPRB1981D, BulaevskiiZhETF1984, AlexandrovPRB1986, AleksandrovZhETF1987, AlexandrovPRB1996, ZhaoPRB1987, MicnasRMP1990, RobaszkiewiczAPPA1994, MicnasPRB1992, MicnasPRB1995, RobaszkiewiczAPPA1996, MicnasCMP1998, PawlowskiPhysA2001}.
Its Hamiltonian has {the} form
\begin{equation}\label{eq:hamHCB}
\hat{H}=-\sum_{ij} I_{ij} (\hat{b}_i^{\dagger} \hat{b}_{j} +H.c.) +\sum_{ij} K_{ij} \hat{n}_i^b \hat{n}_j^b - \bar{\mu} \sum_i \hat{n}_i^b,
\end{equation}
where the model parameters appearing in (\ref{eq:hamHCB})  are related to parameters of {the model} (\ref{eq:EHM}) as follows (up to the second term in $t_{ij}/|U|$ expansion, cf. \cite{RobaszkiewiczAPPA1994}): \mbox{$I_{ij}=2t_{ij}^2/|U|$} describes pair hopping, \mbox{$K_{ij}=I_{ij} + W_{ij}$} represents effective intersite boson-boson interactions, \mbox{$\bar{\mu}=\mu +|U|/2 -W_0$}, 
\mbox{$W_0 = \sum_j W_{ij}$}, and \mbox{$I_0 = \sum_j I_{ij}$}.
Formally, in (\ref{eq:hamHCB}) also the constant term $C= - L \left( I_0 + 2W_0 \right)/4$ appears.
The hard-core boson operators are related to original fermionic ones as {follows:}
\mbox{$\hat{b}_i = \hat{c}_{i\downarrow} \hat{c}_{i\uparrow}$}, 
\mbox{$\hat{b}_{i}^{\dagger}= \hat{c}_{i\uparrow}^{\dagger} \hat{c}_{i\downarrow}^{\dagger}$}, and 
\mbox{$\hat{n}_i^b = \hat{b}_i^\dagger \hat{b}_i$}; {here,} $\hat{b}_i$ and $\hat{b}_i^\dagger$  are commuting (Bose-like) operators for different sites, but {satisfy} \mbox{$\left( \hat{b}_{i} \right)^{2} =\left(\hat{b}_{i}^{\dagger}\right)^2 = 0$} and \mbox{$\left[ \hat{b}_i, \hat{b}_i^{\dagger} \right] = 1 - 2 \hat{n}_{i}^{b}$}.
%In the subspace excluding single occupancy 
{The operators} $\hat{\vec{\rho}}_i$, defined as 
\mbox{$\hat{\rho}^{+}_{i} = \hat{b}_i^{\dagger}$}, 
\mbox{$\hat{\rho}^{-}_i \equiv \left(\hat{\rho}_i^+ \right)^{\dagger} = \hat{b}_i$} and 
\mbox{$\hat{\rho}_i^z = 1/2 - \hat{n}_i^b$}, fulfill commutation relations of spin-$1/2$ operators. 
Thus, {the} hamiltonian (\ref{eq:hamHCB}) can be written in the language of {the} pseudospin operators $\hat{\vec{\rho}}_i$ and (in the subspace excluding single occupancy of electrons) it has {the} form of an anisotropic Heisenberg model {in an} external field, with {a} fixed magnetization equal to $(n-1)/2$~\cite{RobaszkiewiczPRB1981C, RobaszkiewiczPRB1981D, MicnasRMP1990, RobaszkiewiczAPPA1994, MicnasCMP1998}.
The concentration of hard-core bosons is \mbox{$\tilde{n}=n/2$}.

\begin{figure}
    \centering
    \includegraphics[width=\size]{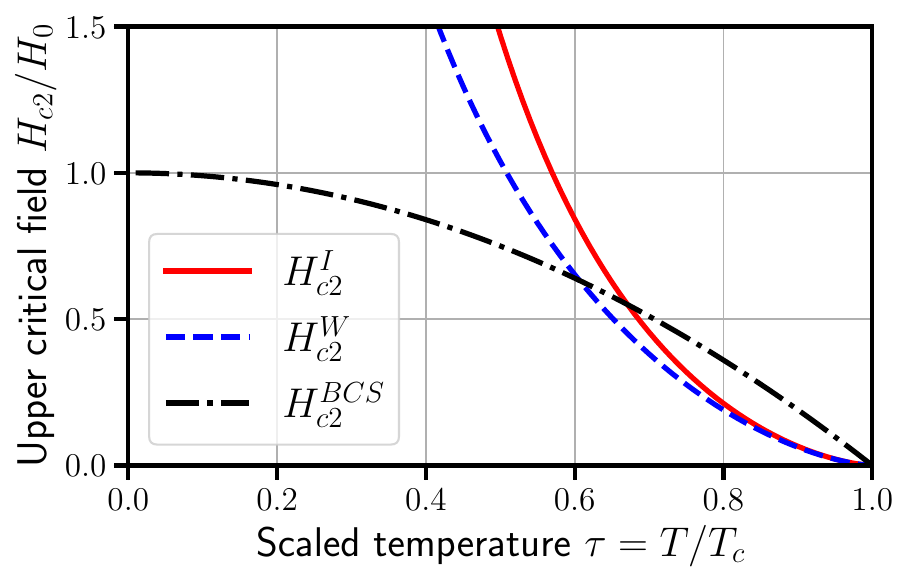}
    \caption{Temperature dependence of {the} upper critical field $H_{c2}$ for hard-core bosons ($H^{\textrm{I}}_{c2}/H^\textrm{I}_0$), weakly interacting bosons ($H_{c2}^{\textrm{W}}/H_0^{\textrm{W}}$), and a second-type BCS superconductor ($H_{c2}^{\textrm{BCS}}/H_{0}^{\textrm{BCS}}$), {(represented by} solid, dashed, and dash-dotted lines, respectively). 
    Expressions given in text.}
    \label{fig:EffectsofH}
\end{figure}

The  phase diagram of model {the}  (\ref{eq:hamHCB}) obtained with the mean-field approximation (MFA) {and} presented in Fig. \ref{fig:PD-HCB-vsn}, has the similar structure as that of {the model} (\ref{eq:EHM}) in the weak-coupling limit discussed {in the previous section.} 
For \mbox{$K/I>1$}  (\mbox{$W>0$}), {the diagram} consists of four regions of the CO, M, SC, and NO phases mentioned previously, with continuous transitions between them \cite{RobaszkiewiczPRB1981C, RobaszkiewiczPRB1981D}.
If \mbox{$K/I=1$}  (\mbox{$W=0$}), only the SC and NO phases appear on the diagram with the boundary line determined by \mbox{$k_{B}T_{c} = zI (n-1) \{ \ln[n/(2-n)] \}^{-1}$} (this boundary is not affected by \mbox{$W \neq 0$}).
At \mbox{$T=0$}, the M phase is degenerated with the PS state, which is the CO and SC phases coexistence \cite{RobaszkiewiczAPPA1996,KapciaJSNM2013}. 
The NNN repulsion \mbox{$W_2>0$} stabilizes the M phase, whereas \mbox{$W_2<0$} eliminates the M phase and extends the stability regions of the PS state at the cost of the homogeneous SC phase.
{It also leads to} the appearance of {a} new PS state being {the} coexistence of the CO and NO phases (for large enough $W$) \cite{MicnasPRB1995, RobaszkiewiczAPPA1996}. 
The existence of the PS states is signalized in \cite{RobaszkiewiczPRB1981D} by {the} nonmonotonous behavior of {the} $n$-dependence of $\mu$ in the homogeneous phases {(the function} $\mu(n)$ should {be increasing one}; \mbox{$\partial \mu/ \partial n >0 $} for homogeneous phases {that are} stable against phase separation).

The most prominent feature of {the} local-pair model (\ref{eq:hamHCB}) is a positive curvature of {the} $T$-dependence of {the} upper critical field \mbox{$H_{c2}(T)=\phi_0 /(2\pi \xi(T)) $} \cite{BulaevskiiZhETF1984, AlexandrovPRB1986, AleksandrovZhETF1987, MicnasRMP1990, AlexandrovPRB1996, DomanskiPRB2003}.
For weakly-interacting bosons (particle-particle scattering included) it was found as $H_{c2}^{\textrm{W}}(\tau) = H_0^{\textrm{W}} \tau^{-1} \left( 1 - \tau^{3/2}\right)^{3/2} $ \cite{AleksandrovZhETF1987}, whereas including impurity scattering one gets
\mbox{$H^{\textrm{I}}_{c2}(\tau) = H^\textrm{I}_0 \tau^{(-3/2)} \left( 1 - \tau^{3/2}\right)^{3/2}$} \cite{AlexandrovPRB1986}, where \mbox{$\tau = T/T_{c}$}. 
Near $T_c$ (\mbox{$\tau \approx 1$}), in both cases, {the same functional dependence \mbox{$H_{c2} \sim \left( 1 - \tau^{3/2}\right)^{3/2} $} describes the nonlinear behavior, which} is a characteristic feature of charged bosons and distinct from the BCS prediction of \mbox{$H_{c2}^{\textrm{BCS}}(\tau) = H^{\textrm{BCS}}_0 \left( 1 - \tau^2 \right)$} for second-type superconductors.  
With decreasing $\tau$ expressions for $H_{c2}^{\textrm{W}}(\tau)$ and $H^{\textrm{I}}_{c2}(\tau)$ diverge, {however, this is a consequence of} the assumptions in the derivations.   
Low-$T$ analysis obtained for both types of scattering predicts finite but high values of \mbox{$H_{c2}\sim T^{-\nu}$} with $\nu$ {ranging} from $0.5$ up to $4.5$~\cite{AlexandrovPRB1996}.
Figure \ref{fig:EffectsofH} presents {the aforementioned} temperature dependence of $H_{c2}$.
Note that {the} early prediction of {a} linear dependence \mbox{$H^{*}_{c2}(\tau) = H_0^*(1-\tau)$} in the vicinity of $T_{c}$ together with the conclusion that superconductivity could not be suppressed by the field for \mbox{$\tau \lesssim 0.88$} was an artifact of the MFA~\cite{BulaevskiiZhETF1984}.

The self-consistent RPA has been used also for the study of model (\ref{eq:hamHCB}) to provide reliable results for  simple cubic lattices with anisotropic nearest-neighbor interactions $R_{ij}$ {(where} \mbox{$R \in \{ I, K \} $}) ($R_{||}$ in $xy$-plane, and $R_{\perp}$ in $z$-direction) \cite{MicnasPRB1992}. 
Such formulation allows {us} to study quasi-2D case (the layered structure, \mbox{$R_{||}  \geq R_{\perp}$}) and quasi-1D case (\mbox{$R_{||}  \leq R_{\perp}$}). 
The limiting cases, {such as an} isotropic 3D simple cubic lattice (\mbox{$R_{||} = R_{\perp}$}), {a} 2D square lattice (\mbox{$R_{\perp}=0$}), and {a} 1D linear chain (\mbox{$R_{||}=0$}) are also included \cite{MicnasPRB1992, MicnasPRB1995}.
Increasing {the} lattice anisotropy {(i.e., decreasing the} interplanar $R_{\perp}$ or interchain $R_{||}$ interactions) extends {the} regions of the SC phase and reduces the CO occurrence {(leading to the} suppression of the M phase).
Due to quantum fluctuations, the boundary between the SC and M phases is effected and the SC phase exists even for \mbox{$K/J \gg 1$} limit if $|x|>|x_c|\neq 1$ \cite{AlexandrovPRB1986, MicnasPRB1992}.
{This result differs from the MFA result,} which gives $|x_c|\rightarrow 1$ in this limit \cite{RobaszkiewiczPRB1981C}.
At finite temperatures the RPA does not predict long-range SC ordering in 1D and 2D cases, in agreement with the exact theorem \cite{MerminPRL1966}.
Moreover, for the system dimensionality {in the range \mbox{$2<d \leq 3$}, it has been established that the} temperature dependencies of the SC phase properties (order parameter, internal energy,  specific heat, London penetration depth, and lower critical field) were found to {follow} power laws {of the form} $AT^{\alpha}$ ($A$ is a numerical factor), both at low and high temperatures (in the latter case, only beyond the critical regime and for low concentrations) with $\alpha$ depending on {the} dimensionality and the temperature regime~\cite{MicnasPRB1992, MicnasPRB1995}. 

Unscreened (long-range) Coulomb interactions $K_{ij}$ (which are a result of electric charge of the particles)  modify the low-$T$ behavior of {the aforementioned quantities, leading to the dependence of the form} $AT^{\alpha} \exp \left[-\Delta/(k_{B}T)\right]$ for low density and 3D {systems} with different exponents $\alpha$ than in the case of only (short-range) NN $K$ interactions~\cite{MicnasPRB1995}.
In the SC phase, the sound-wave-like mode (appearing for short-range interaction) is turned into a plasma-like mode in the presence of unscreened interactions \cite{KostyrkoPRB1992, MicnasPRB1995}.
For long-range $K_{ij}$ repulsion, the collective excitation spectrum  is gapless {for 2D system, whereas in 3D system} it exhibits a gap (cf. also the discussion at the end of Sect.~\ref{sec:weakcopuling}).

\subsection{The model with intersite pairing ($W<0$)}
\label{sec:EHMintersite}

{The} model (\ref{eq:EHM}) with \mbox{$U>0$} and \mbox{$W<0$} incorporates both magnetic and superconducting correlations due to repulsive $U$ and attractive $W$, respectively. 
{Applying the} HFA for {the} model (\ref{eq:EHM}) in this case allows to {derive} expressions for the AF phase (spin-density-wave state), the CO phase (charge-density-wave state) and the SC phases including onsite $s$-wave, extended (intersite) $s$-wave, $d$-wave, and triplet (equal-spin) $p$-wave pairing \cite{MicnasJPC1988,MicnasPRB1988, MicnasPRB1989}.
It was shown that {a} nearest-neighbor attraction $W$ can stabilize various anisotropic superconducting states depending on lattice geometry and band filling.

\begin{figure}
    \centering
    \includegraphics[width=\size]{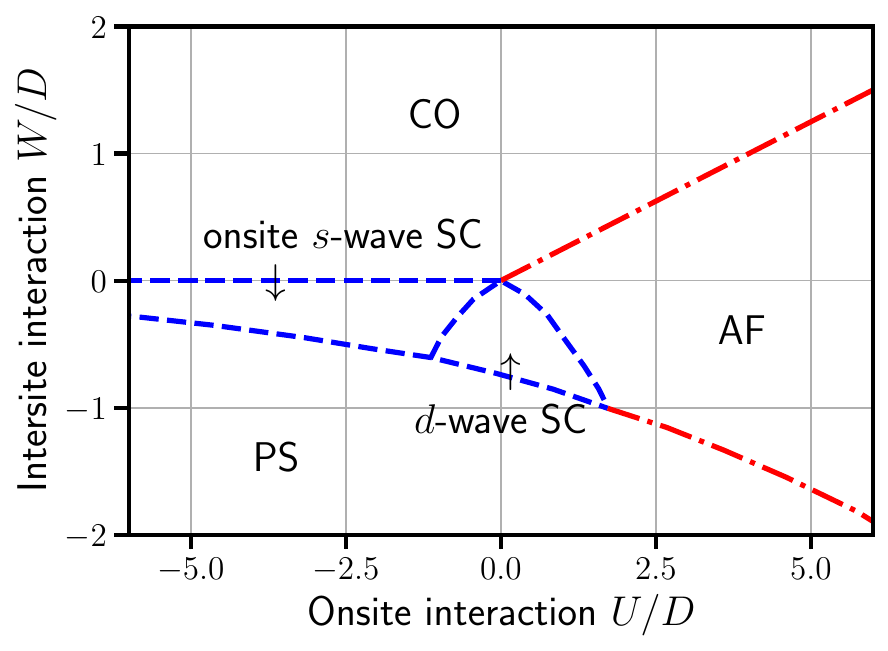}
    \caption{Phase diagram of {the} model (\ref{eq:EHM}) at \mbox{$T=0$} and \mbox{$n=1$} for the 2D square lattice. 
    Data {taken} from Fig.~2 in~\cite{MicnasCMP1998}, cf. also Fig.~2 in~\cite{MicnasPRB1988}.}
    \label{fig:PD-GS-EHM-n1}
\end{figure}

\begin{figure}
    \centering
    \includegraphics[width=\size]{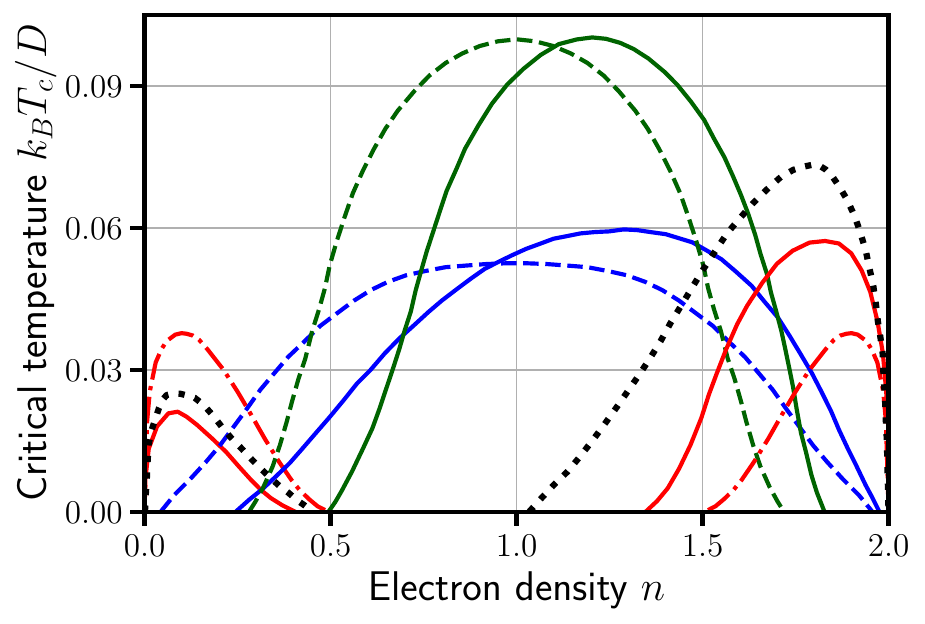}
    \caption{Critical temperature $T_{c}$ for different SC pairings in {the} model (\ref{eq:EHM}) with \mbox{$U=0$}, \mbox{$W/D=-0.5$} on the 2D square lattice (\mbox{$D=4t$}, the HFA result).
    Colors indicate type of SC pairing (red, green, blue for $s$-, $d$-, $p$-wave pairing, respectively).
    The solid curves {refer to the case} of \mbox{$t_2/t =0.3$}, {while}
    dash-dotted and dashed lines corresponds to {the} case of \mbox{$t_{2}=0$}. 
    The dotted (black) line presents {the} results for $s$-wave pairing and \mbox{${\bar{K}}/D=0.25$}.
    Data {taken} from Fig.~8 of~\cite{MicnasPRB1988} and Fig.~7 of~\cite{MicnasPRB1989}.}
    \label{fig:TcWnegvsn}
\end{figure}

The phase diagram for \mbox{$n=1$} is shown in Fig. \ref{fig:PD-GS-EHM-n1} (cf. Refs. \cite{MicnasJPC1988, MicnasPRB1988, DagottoPRB1994, NazarenkoPRB1996, MicnasCMP1998}).
For \mbox{$W>0$}, only two phases are stable, {namely} the CO and AF phases separated by the line \mbox{$U=4W$}.
For sufficiently large $|W|$ (and \mbox{$W<0$}), the PS state (coexistence of two NO domains with different particle concentrations) occurs for any {value of} $U$.
For small and intermediate values of $|W|$, regions of the $s$-wave SC phase (for \mbox{$U<0$}) and of the AF phase {(for \mbox{$U>0$})} exist.
In the center of the diagram (for small negative $W$ and for $U$ near zero), the $d$-wave SC phase is stable.

\begin{figure}[b!]
    \centering
    \includegraphics[width=\size]{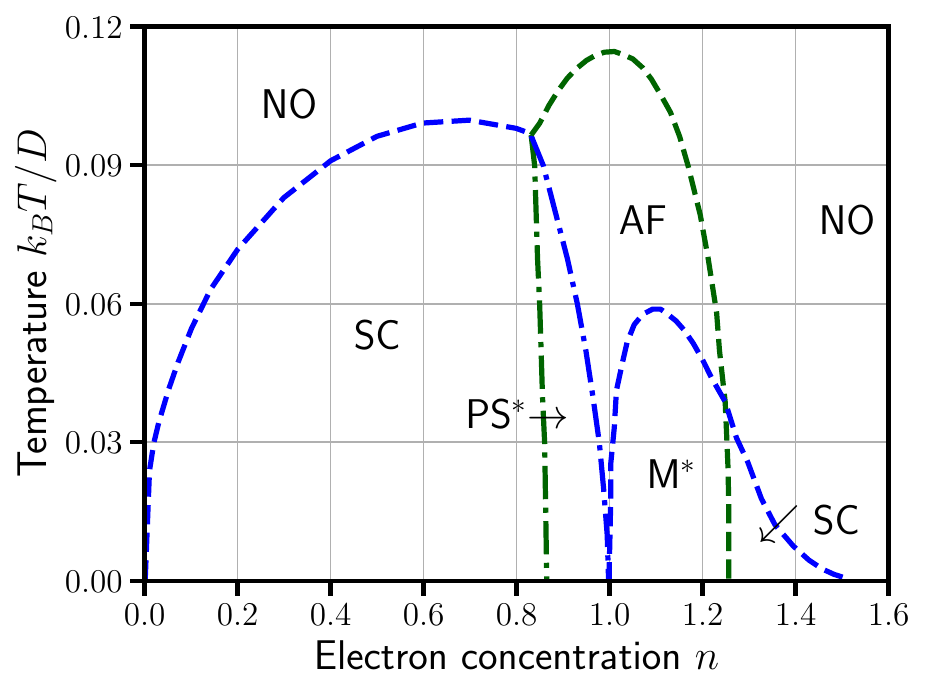}
    \caption{Phase diagram for {the} model (\ref{eq:EHM}) with \mbox{$t_2/t=-0.45$}, \mbox{$U/D=0.9$}, \mbox{$W/D=-0.5$} on the 2D square lattice (\mbox{$D=4t$}, the HFA result).
    All transitions between homogeneous phases are continuous (dashed lines).
    Data {taken} from Fig.~2 in~\cite{TobijaszewskaPSSB2005}.}
    \label{fig:PDWrepAattracvsm}
\end{figure}

The most important insight from the study of {the} model (\ref{eq:EHM}) with negative $W$ is {the} evolution of SC pairing with the doping \mbox{$x = n-1$}.
The relative stability of various superconducting phases was determined by calculating $T_{c}$ for each pairing \cite{MicnasJPC1988, MicnasPRB1988, MicnasJPC1988}.
For arbitrary density, the nature of pairing is strongly dependent on {the} particle concentration.
Figure~\ref{fig:TcWnegvsn} presents the concentration dependencies of $T_{c}$ for the mentioned SC types of pairing.
$d$-wave pairing exhibits highest $T_{c}$ near \mbox{$x=0$}, the competing $p$- and extended $s$-wave parings are stabilized for {larger values of} $|x|$.
Whereas $T_{c}$'s for $d$- and $p$-wave pairings behave \mbox{monotonically} and decrease with {increasing} $|x|$ (or, equivalently, {deviate from the \mbox{$n=1$} case}), $T_{c}$ for $s$-wave pairing {exhibits a non-monotonic dependence} as a function of $|x|$. 
It rises rapidly with decreasing $|x|$ from $1$, goes through the maximum and vanishes continuously to $0$ below some value of $|x|$.
For sufficiently large $|W|$, the $T_{c}$ goes to a finite value for \mbox{$x=0$} (\mbox{$n=1$}).
Note that increasing $U$ reduces $s$-wave pairing and shifts its $T_{c}$ towards higher values of $|x|$.
These results {imply} that the following sequences with increasing $|x|$ are expected: 
(i) $d$--$p$--$s$ for \mbox{$U>0$} or small negative {values of $U$;} 
(ii) $d$--$s$, or only $s$-wave state is {the} most stable for any $n$ when \mbox{$U<0$}.
With increasing $W$, the range of $x$ for $d$-wave SC extends, whereas {the range} for $s$-wave SC shrinks, {while that} for $p$-wave SC slightly shifts towards higher $|x|$.
Note that {the N\'eel temperature $T_{N}$, characterizing} the onset of the AF phase, is rather quickly destroyed {upon} deviating from the \mbox{$n=1$} case (and \mbox{$U>0$}) \cite{TobijaszewskaPSSB2005}.

Inclusion of the next-nearest hopping $t_{2}$ breaks {the} perfect nesting condition and the electron-hole symmetry of {the} model (\ref{eq:EHM}) is not satisfied any longer \cite{MicnasPRB1988, MicnasJPC1988, TobijaszewskaPSSB2005}.
Particularly, for {a} 2D square lattice, the logarithmic singularity in the DOS does not coincide with the Fermi level at \mbox{$n=1$}, and
the maximal $T_{c}$ for $d$- and $p$-wave SC pairing moves to \mbox{$n\neq 1$} \cite{MicnasPRB1988, MicnasJPC1988}. 
The effects of \mbox{$t_{2} > 0 $} are also presented in Fig. \ref{fig:TcWnegvsn}.
Note that $T_{c}$ is substantially changed for small $t_{2}$ and both signs of $t_{2}$ are important, as {the} cases of \mbox{$t_2>0$} and \mbox{$t_2<0$} are related by the relation \mbox{$T_{c}\left( n, t, t_2\right) = T_{c} \left( 2-n,t, - t_{2} \right)$}.
Increasing $t_{2}$ extends the $p$-wave SC region in the under-doped region (i.e., \mbox{$n<1$}), whereas {the} $s$-wave and $p$-wave SC regions are reduced in {the} over-doped regime (i.e., \mbox{$n>1$}).

Attractive {interaction} $W$ may lead to PS states competing with {the} intersite SC pairing mechanism. 
However, {the} inclusion of  repulsive  interactions $W_{ij}$ beyond {the} NN attraction shifts the PS state occurrence region {towards} higher values of $|W|/D$ upon increasing the repulsive longer-range interaction \cite{MicnasPRB1989}.
The NNN {interaction} $W_2>0$  (if $W<0$) can also stabilize the SC phase with $d$-wave pairing suppressing the region of $s$-wave pairing~\cite{MicnasPRB1989, BlaerPRB1997, MicnasCMP1998}.

The effects of phase fluctuations on anisotropic superconductivity of extended $s$- and $d$-wave types were also examined, including {a} time-reversal breaking state of \mbox{$s + \mathbf{i}d$} symmetry  \cite{MicnasJPCM2002, TobijaszewskaAPPA2009}, {as well as} addressing the coexistence of the AF and $d$-wave SC {orderings \cite{TobijaszewskaPSSB2005, SchollePRB2026} and the}  mutual stability of various SC orderings~\cite{TobijaszewskaAPPA2000, TobijaszewskaPSSB2006}. 
The representative finite temperature phase diagram of model (\ref{eq:EHM}) for repulsive $U$ and \mbox{$W<0$} (with \mbox{$t_{2}<0$}) is presented in Fig.~\ref{fig:PDWrepAattracvsm}.
On the diagram, the SC and AF orders can coexist in a~PS state (denoted as PS$^*$) or in a homogeneous mixed state~(indicated by~M$^*$).

\subsection{Other intersite short-range interactions}
\label{sec:other}

In addition to {the} on-site $U$ and intersite $W$ interactions, several other intersite interaction terms, omitted in (\ref{eq:EHM}), can be relevant for real narrow-band systems \cite{MicnasPRB1989, MicnasRMP1990, MicnasCMP1998, KapciaAPPA2012}. 
{These terms} are given by
\begin{eqnarray}\label{eq:EHMadd}
    \hat{H}_{\star} & = & 
    -  \sum_{{i,j}} \bar{I}_{ij} \left( \hat{\rho}^{+}_i \hat{\rho}^{-}_j + \hat{\rho}^{+}_j \hat{\rho}^{-}_i \right)
    -  \sum_{i,j} {\bar{J}_{ij}} \hat{\vec{s}}_i \cdot \hat{\vec{s}}_j \\
    &  & + \sum_{{i,j, \sigma}} \bar{K}_{ij} \hat{n}_{i\bar{\sigma}}\left( \hat{c}^{\dagger}_{i\sigma} \hat{c}_{j\sigma} + \hat{c}^\dagger_{j\sigma} \hat{c}_{i\sigma} \right), \nonumber
\end{eqnarray}
where $\bar{I}_{ij}$, $\bar{J}_{ij}$, and $\bar{K}_{ij}$  are pair hopping, magnetic exchange, and correlated hopping terms (between $i$ and $j$ sites), {respectively.}
The components of spin operators $\hat{\vec{s}}_i$ are defined as \mbox{$\hat{s}_{i}^{+} = \left( \hat{s}_{i}^{-} \right)^{\dagger} = \hat{c}_{i\uparrow}^{\dagger}\hat{c}_{i\downarrow}^{\ }$} and \mbox{$\hat{s}_{i}^{z} = \left( \hat{n}_{i\uparrow} - \hat{n}_{i\downarrow} \right)/2$}.
Note that in (\ref{eq:EHMadd}), the interaction parameters are not dependent {of the} $t_{ij}/U$ ratio, and the single-site space consists of four states, contrary to the limits \mbox{$U\rightarrow \mp \infty$}, where \mbox{$I_{ij} = 2t_{ij}/|U|$} (single occupancy excluded; two single-site states possible) and \mbox{$J_{ij}=2t_{ij}/U$} (double occupancy excluded; three single-site states possible), respectively.
{These parameters} modify the SC phases stability and may substantially enhance or suppress SC pairing tendencies.
Their influence was systematically analyzed in subsequent studies of narrow-band superconductors, {in which} intersite electron pairing {occurs. These studies} are briefly revised below.

Analysis of the HFA equations presented in Refs. \cite{MicnasPRB1988, MicnasPRB1989}, shows that the exchange NN term $\bar{J}$ modifies the pairing potentials for the SC states {as follows:} \mbox{$|W|\rightarrow |W|+3\bar{J}/2$} for extended $s$- and $d$-wave and \mbox{$|W|\rightarrow |W|-\bar{J}/2$} for $p$-wave states, whereas the AF state is modified {accordingly} by \mbox{$U\rightarrow U + z\bar{J}$}.
Therefore, both the $s$- and $d$-wave pairing are enhanced by antiferromagnetic {exchange (\mbox{$\bar{J}>0$}),} whereas the $p$-wave SC phase is suppressed (ferromagnetic $\bar{J}$ {has the opposite effect).}
However,  the mutual stability between the $s$- and $d$-wave SC states is also {modified, namely}  \mbox{$\bar{J}>0$} extends {the} electron concentration range for the $d$-wave state for sufficiently large $\bar{J}$, with nonmonotonous behavior depending on $U$ for small $\bar{J}$.

Regarding {the $\bar{K}$} term {(restricted to NN case)}, at the HFA level, it modifies $s$-wave pairing essentially, whereas other pairings are almost unaffected \cite{MicnasPRB1989}. 
As a result, {\mbox{$\bar{K}>0$}} increases (decreases) $T_{c}$ for  {the} $s$-wave SC phase in {the} overdoped (underdoped) region, influencing the relative stability of the SC phases (cf. {the} dotted line in Fig. \ref{fig:TcWnegvsn}).
Moreover, the $s$-wave SC state {can} be stabilized solely by the {$\bar{K}$} term only, even {when} other mechanism of pairing (i.e., \mbox{$U<0$}, \mbox{$W<0$}, or \mbox{$J\neq 0$}) are absent.

The pair-hopping {NN} term $\bar{I}$ can induce superconductivity independently of the remaining interactions included in (\ref{eq:EHM}) and (\ref{eq:EHMadd}). 
{The case of} \mbox{$\bar{I} > 0$} favors the $s$-wave pairing, whereas \mbox{${\bar{I}} < 0$} stabilizes the $\eta$-wave state.
A model containing only single-particle hopping $t_{ij}$ and pair-hopping $\bar{I}_{ij}$  interaction {(assuming} \mbox{$U=0$}) is known as the  Penson-Kolb (PK) model.
It was originally introduced  in \cite{PensonPRB1986, Kolb1986} and subsequently studied extensively for both signs of {the} $\bar{I}$ interaction, see, e.g., \cite{RoyPRB1997, CzartPRB2001, RobaszkiewiczPSSB2003, MierzejewskiPRB2004, CzartAPPA2004, CzartPSSB2007,  PtokAPPA2008, CzartPhysA2022}.
The model provides a nonlocal pairing mechanism through {the} intersite $\bar{I}$ term, in contrast to the local attraction \mbox{$U<0$} in {the} model (\ref{eq:EHM}). 
In the PK model, {the} $\bar{I}$ term is responsible for both pair formation and also of their condensation.
An important difference between the negative-$U$ Hubbard model (\mbox{$W=0$}) and the PK model is present for \mbox{$n=1$}.
The latter posses $U(1)$ symmetry of the charge channel  for any $n$, whereas the attractive Hubbard model exhibits pseudospin $SU(2)$ symmetry only at \mbox{$n=1$}.
Therefore, in 2D case and at \mbox{$n=1$}, there is no a KT transition in the Hubbard model, whereas for the PK model, the KT critical temperature reaches maximum \cite{CzartPRB2001}.
Introducing \mbox{$U \neq 0$} (the PK-Hubbard (PKH) model) leads to {the} appearance of the AF and CO states as well as various bond orderings in low dimensions \cite{HuiPRB1993, JaparidzeJPCM1997, RobaszkiewiczPRB1999, JaparidzePRB2001, KapciaCzart2016, KapciaCzart2018, CzartJMMM2019}. 
However, a detailed discussion of {these} models is beyond the scope of this review.
It is worth to mentioning that this extension of the PK model can also support {the} existence of the SC states with nonzero center-of-mass momentum of Cooper pairs, including the Fulde-Ferrell and Larkin-Ovchinnkov (FFLO) phases, cf. \cite{CasalbouniRMP, PtokJPCM2009, MierzejewskiPRB2009, PtokPRB2011, PtokSST2015, KapciaJPSJ2016}.

The atomic limit (\mbox{$t_{ij}=0$}) of the PKH model provides a simple framework for studying the mutual competition between intersite pair hopping $\bar{I}$ and onsite repulsion \mbox{$U>0$} in the SC phase \cite{RobaszkiewiczPhysC, KapciaJPCM2012, KapciaJPCM2013, KapciaJSNM2014, KapciaAPPA2014}.
It can also be viewed as a generalization of the hard-core boson model (\ref{eq:hamHCB}) with \mbox{$K_{ij}=0$} to the case of finite pair-binding energy $E_b$.  
Two distinct regimes can be identified. 
In the local pair limit, i.e., for \mbox{$E_{b}/(k_{B}T_{c})\gg 1$}, the SC transition occurs through the Bose condensation of preformed pairs. 
In contrast, for \mbox{$E_{b}/(k_{B}T_{c}) \lesssim 1$}, the transition is governed by pair-breaking excitations and preformed pairs are essentially absent above $T_{c}$.
The crossover between these two regimes, driven by increasing $U$, is continuous.

\section{Experimental scope of the theory}
\label{sec:realsystems}

\subsection{Solid state examples}\label{sec:solidstate}

The theory reviewed above can be applied to {described} many, but by no means all, features of a number of superconducting narrow-band systems.
The interplay of the on-site and intersite attractive or repulsive interactions included in the Hamiltonians leads to competition or cooperation of various electron orderings: superconductivity, charge orderings or magnetism.
It is therefore natural to seek experimental exemplifications of the theory among materials {in which} the mentioned orderings coexist or compete depending on thermodynamic conditions, electron or hole concentration or applied external fields.

{\em The cuprates}. 
Concerning the copper-oxide superconductors, the accumulated body of research indicates that these systems differ substantially from {typical} metals both in the normal phase and from BCS-type superconductors in the ordered phase. 
In particular, the linear scaling of the superconducting $T_{c}$ with the inverse square of the London's penetration depth, measured {in the} \mbox{$T=0$}~K limit, was found for all values of the hole concentrations, except near the overdoping limit \cite{BozovicNature2016,BozovicPhysA2019}, where these systems are more like the conventional BCS-type superconductors. 
Above $T_{c}$ the existence of the preformed electron pairs up to the doping dependent temperature \mbox{$T^{\star} >T_{c}$} was commonly confirmed.  
These facts, together with the very short coherent length in the superconducting phase in cuprates, are widely accepted as corroborating a theory assuming the local electron pairing in these systems.

Relatively recent experimental findings confirmed {the} coexistence of both superconductivity and charge orderings in a wide range of hole doping and temperatures \cite{Keimer2015}.  
Charge orderings in these systems, most often referred to as stripe phases, are usually explained by {the} competition {between the tendency toward} antiferromagnetic ordering due to strong on-site Hubbard repulsion and {the gain in} the electron kinetic energy in the broken symmetry charge state \cite{WhitePRL1998,ScalapinoPhysC2012}.

The chief magnetic properties in the underdoped phase of the cuprates and the $d$-wave symmetry of the electron pairing in the superconducting phase in the optimally doped region clearly require the on-site repulsive interaction. 
In this case, {a} version of the model with the intersite electron attraction (even if treated only at the mean-field level of approximation) can semiquantitatively describe these systems. 
It seems significant that a wave vector defining the charge-density wave varies with the {hole-doping level} as described by the simple $t$-$J$ model in the La-Ba-Cu-O system, with {a} completely opposite tendency in Y-Ba-Cu-O systems \cite{Keimer2015}. 
The difference {has not yet been} explained and it is tempting to believe that it may be due to a different role played by the longer range Coulomb interactions - neglected in the $t$-$J$ Hamiltonian - {across} the two groups of cuprates.

{\em The BaBiK oxide}. 
One of the systems long time ago suggested as a representative example of a superconductor with local electron pairing was potasioum-doped BaBiO$_3$ perovskite, {in which} Bi$^{4+}$$\to$Bi$^{5+}+$Bi$^{3+}$ charge disproportionation {occurs.} 
In this system the transition to {a} superconducting state can {take place at} \mbox{$T_{c}\sim 34$}~K depending on the doping, a value which was earlier considered {to be} above the limit of the conventional electron-phonon coupling mechanism. 
Moreover, the system exhibits a charge ordered state for {lower} values of K-doping. 
It was therefore considered a promising candidate for a description in terms of the model with on-site Hubbard attraction and a resulting $s$-type pairing. 
Recent experiments \cite{GriffittNatCom2023}, however, cast doubts {on} this interpretation: the diffuse X-ray scattering does not show neither long nor short range site charge disproportionations in the superconducting phase, contrary to what should be expected based on the basic negative-$U$ Hubbard model. 
A~still more recent study~\cite{MenushenkovPRR2024} using X-ray absorption spectroscopy concludes that a local pair-density wave consisting {of an} alternation of hole pairs and electron pairs in the neighboring oxygen octahedra may take place in this compound.

{\em The chalcogenides.}  
A tendency toward charge disproportionation (``valence skipping'') was also found in some chalcogenide systems. 
In the AgSnSe$_2$ compound Sn$^{2+}$ and Sn$^{4+}$ ions were shown to coexist using X-ray studies \cite{KatariaPRB2023}, {suggesting that} negative on-site attraction {may provide} a possible explanation { for this.} 
The compound exhibits superconductivity with \mbox{$T_{c}=4.91$}~K and $s$-type symmetry of the pairing function. 
The position of the compound in the Uemura plot and the value of the gap to $T_{c}$ ratio situates it in the region of
moderate to strong coupling. 
Another example of {a} similar type of charge disproportionation is provided by Tl-doped PbTe. 
Some indications of superconductivity due to boson-like electron pairs come from the study \cite{NaskarPRB2022} of La$_3$Se$_4$ material, which exhibits a positive  curvature of $H_{c2}(T)$ near its \mbox{$T_c=8.5$}~K, well fitted by the formula obtained for charged boson systems \cite{AlexandrovPRB1986, AleksandrovZhETF1987, MicnasRMP1990}.
Of particular interest is superconductivity in {the} ZrTe$_3$ compound \cite{TsuchiyaNJP2017}, which exhibits high anisotropy
of electron transport and {has a critical temperature of} \mbox{$T_{c}\sim4$}~K; {superconductivity exists} well inside a charge-density-wave state. 
It was argued that the coherence length in this system is extremely small due to almost flat electron band near the
Fermi level, the former feature suggesting strongly local electron pairs.

{\em Magn\'eli phases.} 
The titanium oxides with varying content of oxygen and a unique spatial structure based on TiO$_2$ slabs were already considered {in the 1970s of the XX century} as systems with preformed intersite electron pairs existing in the normal
state \cite{MarezioJSSC1973,LakkisPRB1976}. 
Such {an} interpretation was invoked to explain specific changes in transport and magnetic susceptibility with change of temperature in these systems. 
However, only quite recently a superconducting state in these systems was found \cite{YoshimatsuSciRep2017}.
The critical temperature as high as 7~K was reached in {the} Ti$_3$O$_5$ compound in epitaxial thin films.  
Although it is yet too early to draw a definite conclusion about {the} nature of superconductivity in these systems, one can expect that they may be considered as examples of superconductor with intersite electron pairing, suitable for analysis in terms of a model with intersite electron attraction.

\subsection{Ultracold atoms on optical lattices}\label{sec:ultracold}

Ultracold atoms in optical lattices have transformed the Hubbard model from a theoretical concept into a highly controllable platform for quantum simulation of interacting many-body systems \cite{BlochNatPhys2012, EsslingerAR2010, JakschAP2005, DuttaRPP2015, GeorgescuRMP2014}. 
Precise control of lattice geometry, interactions, and trapping potentials enables clean studies of phenomena such as magnetism and superconductivity, while providing direct links between theory and experiment \cite{BlochRMP2008, HartNat2015}. 
Consequently, reliable theoretical and numerical methods are essential for interpreting experimental data and benchmarking quantum-simulation platforms.

One of the fundamental open questions is whether ultracold Fermi gases can provide insight into the nature of the normal state of strongly interacting fermionic systems above the superfluid transition temperature. 
Understanding this regime may also shed light on the mechanisms responsible for high-$T_{c}$ superconductivity.
On the theoretical side, a major challenge is to develop a quantitative description of the BCS-BEC crossover that consistently accounts for both the normal and superfluid phases. 

The BCS-BEC crossover has been extensively explored experimentally in ultracold two-component Fermi gases, where the interaction strength can be continuously tuned across a Feshbach resonance. 
Experiments on strongly interacting $^{6}\mathrm{Li}$ and $^{40}\mathrm{K}$ gases have demonstrated superfluidity throughout the crossover and provided direct access to the evolution of pairing and single-particle excitations from the BCS regime to the BEC regime \cite{ZwierleinNat2006, StewartNat2008}. 
Momentum-resolved photoemission and radio-frequency spectroscopy have further revealed an energy gap that persists above the superfluid transition.
{This proves} experimental evidence for a pseudogap regime in which fermionic pairs exist without long-range phase coherence \cite{GaeblerNP2010, FeldNat2011}. 
More recently, momentum-resolved microwave spectroscopy of homogeneous unitary $^{6}\mathrm{Li}$ gases has provided quantitative evidence that the pseudogap can be driven by pairing fluctuations above $T_{c}$ \cite{LiNat2024}. 
These experiments establish ultracold Fermi gases as a particularly clean, unique and highly controllable platform for investigating the relationship between pair formation, phase coherence, and the normal state of strongly interacting fermionic matter.
 
An unconventional manifestation of pairing and phase coherence is spatially homogeneous spin-polarized superconductivity (breached pair), which is characterized by a gapless spectrum for the majority spin species~\cite{KujawaAPPA2008, KujawaEPL2011, CichyAP2014, CichyAP2015, PtokPRA2017, PtokJSNM2018, CichyPRB2022, CichyJMMM2024}.
Spin imbalance and synthetic gauge fields have enabled the exploration of a variety of exotic superfluid states \cite{RadzihovskyRPP2010}. 
Experiments with imbalanced $^{6}\mathrm{Li}$ gases have mapped the competition between paired and polarized phases, revealing first-order superfluid-to-normal transitions and tricritical behaviour \cite{ShinNat2008, GuoPRA2009}. 
In lower-dimensional geometries, spin-imbalanced Fermi gases exhibit partially polarized phases consistent with the physics expected for {FFLO-type} pairing \cite{LiaoNat2010}. 
In addition, the realization of synthetic spin–orbit coupling in ultracold Fermi gases has opened {the} experimental routes toward topological and other unconventional superfluid states \cite{HuangNatPhys2016, Torma, Melo, Wilczek, Gubankova, Iskin, Parish}.
{Taken together,} these developments demonstrate that ultracold gases can reproduce and experimentally probe pairing phenomena that are difficult to isolate in conventional superconducting materials.

It is worth emphasizing that several extensions of the attractive Hubbard model discussed in the previous sections can be realized using ultracold atoms in optical lattices. 
In particular, Bose-Hubbard systems also provide experimental realizations of lattice bosons with local repulsion.
They have enabled detailed studies of the superfluid-Mott insulator transition, finite-temperature phase diagrams, and the effects of artificial gauge fields~\cite{Fisher, Zoller2, Cooper01112008, Goldman, SajnaPRB2018, SAJNA2018215, SAJNA2019257}.

These developments are particularly important because they provide experimentally controllable realizations of many-body Hamiltonians closely related to the extended Hubbard {models} and hard-core boson models discussed in this review.
As a consequence, ultracold atomic systems have become an important testing ground for theoretical concepts such as local pairing, phase fluctuations, pseudogap formation, and the BCS-BEC crossover.

\section{Summary and final remarks}
\label{sec:summary}

In this review we have summarized the development of theoretical models with non-retarded short-range attractive interactions, originally formulated and extensively studied by R. Micnas, S. Robaszkiewicz, and their collaborators. 
These models were motivated by the search for microscopic mechanisms of superconductivity in narrow-band systems and high-$T_{c}$ materials, where local pairing and short coherence lengths may play an essential role.
They can account for the existence of the preformed electron pairs in the normal phases of strange metals, as well as the coexistence of the superconducting state and the charge density waves which are frequently
spotted in many narrow-band materials.

The models have been investigated using a variety of analytical and numerical methods, ranging from mean-field and random-phase approximations to more sophisticated many-body techniques. 
{Such studies have} yielded analytical and numerical results for the ground-state properties, excitation spectra, transition temperatures, electrodynamic response, and collective modes over a broad range of interaction strengths and carrier concentrations. 
In particular, they provided a coherent description of the evolution from weak-coupling BCS superconductivity to the strongly coupled local-pair regime.
A distinctive feature of the approach is that the microscopic origin of the attraction is not specified explicitly. 
Instead, it is assumed that coupling to  bosonic excitations, such as phonons, excitons, or other collective modes, leads to an effective instantaneous attractive interaction between electrons. 
As a consequence, electron pairing occurs predominantly in real space, providing a natural framework for describing local-pair superconductivity, preformed pairs, charge ordering, and the continuous crossover between the BCS and BEC limits.
Notice that, apart from the numerical analyses performed in the works presented in Sect. \ref{sec:EHMgeneral}, many analytical results in various limits were derived there.
Unfortunately, due to space limitation, they cannot be presented and discussed here.

The simplicity and generality of the considered models make them valuable reference systems, {useful both} for understanding strongly correlated superconductors and for benchmarking advanced numerical approaches, such as Monte Carlo {simulations,} DMFT, {or the} density-matrix renormalization group (DMRG) calculations.

The models reviewed here are restricted to an effective single-band description of narrow-band systems. 
Consequently, some material-specific features, such as detailed Fermi-surface properties or isotope effects, lie beyond the scope of this approach.
A natural extension of the local-pair framework is provided by two-component models, {in which} itinerant electrons coexist and interact with local electron pairs, e.g.,  \cite{RanningerPhysBC1985, FriedbergPLA1989, FriedbergPRB1989, MicnasPRB2002, MicnasPhysC2003, DomanskiPRB2003, DomanskiPRL2003, PawlowskiJS2004, MicnasSB2005, Micnas2007, RobaszkiewiczPRB1987, RomanoPRB1997, PawlowskiAPPA1998, CzartAPPA2001, CzartIJMPB2001}. 
In these models, superconductivity emerges through a charge-exchange mechanism between localized bosonic pairs (also {having a} finite pair-binding energy being {an} extension of the standard hard-core boson-fermion model \cite{PawlowskiAPPA1998, CzartAPPA2001, CzartIJMPB2001}) and itinerant fermions, leading to a mutual induction of superconducting correlations in both subsystems.
Such two-component models offer a unified description of local-pair superconductivity, pseudogap phenomena, and the crossover between BCS-like and Bose-condensed regimes. 
The effects of disorder have also been extensively investigated, e.g., {in}~\cite{KrzyszczakJPCM2010,Pawlowski2010}. 
Depending on the relative concentration of local pairs and itinerant carriers, {these models} reproduce a wide variety of behaviors ranging from conventional superconductivity to regimes dominated by preformed pairs and strong phase fluctuations.

Finally, it should be emphasized that {models of superconductivity} with purely short-range interactions are physically incomplete, since they neglect the long-range Coulomb interaction associated with the electric charge of electrons. 
Although such models remain useful for studying phase competition and the physics of local pairing, their applicability to electrodynamic and low-temperature thermodynamic properties is necessarily limited.

\begin{acknowledgments}
We thank our present and former colleagues from {the} Department of Theory of Condensed Matter (AMU, Pozna\'n): Maciej B\k{a}k, Ravindra W. Chhajlany, Wojciech R. Czart, Przemys\l{}aw R. Grzybowski, {Mariusz M. Nogala,} Tomasz P. Polak, Marek Thomas, and Ryszard J.~\mbox{Wojciechowski} for very fruitful discussions and useful comments concerning the topics {addressed in} this review article.%
\end{acknowledgments}

%\bibliography{biblio-SS}


\end{document}